\documentclass[letterpaper,twoside,10pt]{article}
\usepackage{amssymb}
\usepackage{latexsym}
\usepackage{verbatim}
\usepackage{floatfig,epsfig}
\usepackage{calc}
\usepackage{fancyheadings}
\usepackage{pstricks,pst-node}
\usepackage{rotating,graphicx}
\newlength{\figurewidth}
\newcommand{\be}{\begin{equation}}
\newcommand{\ee}{\end{equation}}
\newcommand{\ble}[1]{\begin{equation} \label{#1}}
\newcommand{\bae}{\begin{eqnarray}}
\newcommand{\eae}{\end{eqnarray}}
\newcommand{\fle}[2]%
{\vspace{1.5ex}
\be
\label{#1}
\mbox{%
\setlength{\fboxsep}{3ex}%
\framebox{$\dss #2 $}}
\ee}
\newcommand{\equalref}[1]{\stackrel{\rule[-1ex]{0mm}{0mm}%
{}_{(\ref{#1})}}{=}}
\newcommand{\nn}{\nonumber}
\newcommand{\ff}{\nn \\}
\newcommand{\fe}{& = &}

\newtheorem{exatitle}{Example}[section]
\newenvironment{example}[2]%
{\begin{exatitle} \label{#2} #1 \end{exatitle}}%
{\hfill $\Box$ \\}
\newenvironment{aside}%
{\vspace*{3mm} \noindent {\bf Aside:} \small}%
{\hfill \rule[-3mm]{0mm}{0mm}$\Diamond$\\}

\newcommand{\dss}{\displaystyle}

\newcommand{\Ra}{\Rightarrow}
\newcommand{\lbbr}{[\! [}
\newcommand{\rbbr}{]\! ]}

\newcommand{\tr}{\triangleright}

\newcommand{\eg}{\hbox{\it e.g.{}}}
\newcommand{\etc}{\hbox{\it etc.{}}}
\newcommand{\ie}{\hbox{\it i.e.{}}}
\newcommand{\wrt}{\hbox{w.r.t.{}}}

\newcommand{\rhs}{\hbox{r.h.s.{}}}

\newcommand{\capitem}[1]{\caption{\textsf{#1}}}

\newcommand{\calH}{\mathcal{H}}

\newcommand{\hH}{\hat{\calH}}
\newcommand{\bBb}{\lbbr B \rbbr}
\newcommand{\bnb}{\lbbr n \rbbr}
\newcommand{\hpar}{\partial_q}
\newcommand{\hx}{x_q}

\newcommand{\tf}{\widehat{f}}

\newcommand{\ketn}{| n \rangle}
\newcommand{\ketnm}{| n-1  \rangle}
\newcommand{\ketnp}{| n+1 \rangle}
\newcommand{\ketqn}{| n \rangle_q}
\newcommand{\ketan}{| n \rangle_\alpha}

\newcommand{\vacz}{| 0 \rangle}
\newcommand{\averx}[1]{\langle #1 \rangle_x}
\newcommand{\averqx}[1]{\langle #1 \rangle_x^q}

\newcommand{\papertitle}{%
Canonical Commutation Relation Preserving Maps%
}
\newcommand{\headtitle}{%
\papertitle%
}
\newcommand{\paperauthor}{%
C. Chryssomalakos and A. Turbiner%
}
\begin{document}
\initfloatingfigs
\begin{titlepage}
\vspace*{-1cm}
\begin{flushright}
\textsf{ICN-UNAM-01/06}\\
\textsf{LPT-ORSAY 01-29}
\\
\mbox{}
\\
\textsf{March 31, 2001}
\\[3cm]
\end{flushright}
\renewcommand{\thefootnote}{\fnsymbol{footnote}}
\begin{LARGE}
\bfseries{\sffamily \papertitle}
\end{LARGE}

\noindent \rule{\textwidth}{.6mm}

\vspace*{1.6cm}

\noindent \begin{large}%
\textsf{\bfseries \paperauthor}%
\end{large}%
\footnote{%
Present address (on sabbatical leave): 
Laboratoire de Physique Theorique, 
Universit\'e Paris Sud, 
Orsay 91405, 
France. 
On leave of absence from the
Institute for Theoretical and Experimental Physics, 
Moscow 117259, 
Russia.%
}


\phantom{XX}
\begin{minipage}{.8\textwidth}
\begin{it}
\noindent Instituto de Ciencias Nucleares \\
Universidad Nacional Aut\'onoma de M\'exico\\
Apdo. Postal 70-543, 04510 M\'exico, D.F., MEXICO \\
\end{it}
\texttt{chryss@nuclecu.unam.mx, turbiner@nuclecu.unam.mx
\phantom{X}}
\end{minipage}
\\

\vspace*{3cm}
\noindent \textsc{\large Abstract: } 
We study maps preserving the
Heisenberg commutation relation $ab - ba=1$. We find a
one-parameter deformation of the standard realization of the above
algebra in terms of a coordinate and its dual derivative. It
involves a non-local ``coordinate'' operator
while the dual ``derivative'' is just the Jackson
finite-difference operator. Substitution of
this realization into any differential operator involving $x$ and
$\frac{d}{dx}$, results in an {\em isospectral} deformation of a
continuous differential operator into a finite-difference one. We
extend our results to the deformed Heisenberg algebra $ab-qba=1$.
As an example of potential applications, various deformations of
the Hahn polynomials are briefly discussed.
\end{titlepage}
\setcounter{footnote}{0}
\renewcommand{\thefootnote}{\arabic{footnote}}
\setcounter{page}{2}
%
%
\section{Introduction}
\label{Intro}
The Heisenberg algebra
 \footnote{%
We ocassionally refer to~(\ref{Ha}) in the sequel as the {\em 
canonical commutation
relation} (CCR).} 
\ble{Ha} 
[a, \, b] =1 
\ee 
made its first
appearance in physics, long before the birth of Quantum Mechanics,
through its realization involving a continuous coordinate $x$ and
a dual derivative $d/dx \equiv \partial$, the latter being the
basic differential operator of analysis.
General differential operators, in one dimension, are then
expressed in terms of powers of $\partial$, multiplied by
functions of $x$ --- a wide class of physical problems leads to an
eigenvalue equation for such operators.

The reason underlying the predominance of this particular
realization in physics is the continuous nature of most spaces
under study. Recently, however, there has been a growing interest
in discretized versions of spacetime or other, internal spaces.
This sometimes originates in the need for numerical computation,
as in, \eg, lattice QCD, but has also been proposed as a model of
small scale structure. On another front, alternative realizations
of~(\ref{Ha}) have emerged in string theory (see,
\eg,~\cite{Sch:91}, \cite{Ana.Bow.Sch:92}). In either case, $x$,
$\partial$ ceases to  be the realization of choice and, in several
cases, discrete (finite difference) operators acquire preferred
status. Maintaining the validity of~(\ref{Ha})
makes the transition from the continuous to the discrete
non-trivial. The need then arises for new realizations of the
Heisenberg algebra in terms of discrete operators. Given such
realizations, the differential operators mentioned above can be
deformed by replacing the continuous realization by a discrete
one --- the non-trivial feature of such deformations is that they
are isospectral%
\footnote{%
We do not consider here the effect of boundary conditions.%
}. 
The process may be regarded as a quantum
canonical transformation.

There has been already a considerable amount of research in this
direction (see, \eg,~\cite{Smi.Tur:95, Smi.Tur:96, Gor.Szm:98,
Gor.Szm:00, Tur:00} and references therein), with the discrete
derivative $\partial_\delta$, defined by
\ble{partdel}
 \partial_\delta \tr f(x) \equiv
 \delta^{-1} \bigl( f(x + \delta) -f(x) \bigr) \, ,
\ee
receiving traditionally most of the attention. It can be argued
that this is due, in part, to the fact that the form of the
canonically conjugate ``coordinate'' variable $x_\delta$ is known
(see~\cite{Smi.Tur:95} and Ex.~\ref{fdreal}). It is clear, from its 
definition, that
$\partial_\delta$ can be restricted to the (equally spaced) points
of a lattice. A second natural choice would be an
exponential lattice, the corresponding finite difference operator
being the {\em Jackson derivative} (or {\em Jackson symbol}, 
see~\cite{Ext:83}), defined by
\be
\label{hparactf}
 \hpar \tr f(x) = \frac{1}{1-q} x^{-1}
 \bigl(f(x)-f(qx)\bigr) 
\, . 
\ee
The problem with this choice, and the main motivation for this
work, is that the form of the canonically conjugate ``coordinate''
operator $x_q$ seems to be unknown. We solve this problem in
Sec.{}~\ref{JDCC} below while in Sec.{}~\ref{CCRpMAB} we study
representations, in a very general setting. Sec.{}~\ref{Compo}
shows how to generate new canonical commutation relation
preserving maps from known ones and Sec.{}~\ref{qCC} briefly
extends the above results to the case of $q$-canonical commutation
relations. In Sec.~\ref{Hahn} two isospectral deformations of the
Hahn operator are presented as a concrete application --- the
corresponding polynomial eigenfunctions are also supplied.
\section{The Jackson Derivative and its Canonical Conjugate}
\label{JDCC}
Given the Jackson derivative $\hpar$, satisfying
\be
\label{qccr}
\hpar x -q x \hpar =1 \, , \qquad -1 < q < 1
\, .
\ee
One finds
\be
\label{hparactxn}
 \hpar \tr x^n = \{ n \} x^{n-1} \, , \quad
 \qquad \{n\} \equiv \frac{1-q^n}{1-q} \, , \quad n=0,1,2,\ldots
\ee
 where
\ble{actdef}
 \hpar \tr x^n \equiv \hpar x^n \vacz_{x,\hpar} \, .
\ee
 The notation used in the \rhs \ of the above equation is as
follows. $\vacz$ denotes the ``vacuum'', a ket annihilated by
derivatives, $\hpar \vacz=0$. The subscript of $\vacz$ is an
instruction: express all variables to its left in terms of $x$,
$\hpar$ (already in this form, in this particular example), then
use the commutation relation~(\ref{qccr}) to bring the $\hpar$'s
to the right of the $x$'s. There they are annihilated by $\vacz$
leaving a function of $x$ only --- this function serves to define
the l.h.s., \ie, \ the {\em action} of $\hpar$ on $x^n$. For a
general function $f(x)$, defined as a Taylor series in $x$, the
above relation leads to the alternative
definition~(\ref{hparactf}),
which makes it clear that $\hpar$ acts on the exponential lattice
$\{x, \, qx, \, q^2x, \ldots \}$.
$\hpar$ can be realized as a pseudodifferential operator
\be
\label{hparpseudo}
\hpar \sim \frac{1}{1-q}x^{-1} (1-q^A) \equiv x^{-1} \{ A \}
\, ,
\qquad
\quad
A \equiv x \partial
\, ,
\ee
 where $\partial$ is the partial derivative \wrt \ $x$, $\partial x=1
+ x \partial$, also annihilating the vacuum, $\partial \vacz=0$.
For the reasons mentioned in the introduction, one would like to
realize also an operator $\hx$, such that $\hpar \hx = 1 + \hx
\hpar$. Using the commutation relation $A x^{-1}=x^{-1}(A-1)$ we
find $x^{-1} A^n = (A+1)^{n-1} \partial$, so that ($q \equiv e^h$)
\bae 
\label{hparB} 
\hpar \fe \frac{1}{1-q}x^{-1} (1-q^A) \ff
 \fe - \frac{1}{1-q} x^{-1} \sum_{n=1}^{\infty}
\frac{h^n}{n !} A^n \ff
 \fe - \frac{1}{1-q} x^{-1} \sum_{n=1}^{\infty}
\frac{h^n}{n !} (1+A)^{n-1} \partial \ff
 \fe - \frac{1}{1-q} (1+A)^{-1} (q^{1+A} -1) \partial \ff
\Rightarrow \, \, \hpar \fe \frac{1}{1-q} B^{-1} (1-q^B)
\partial
\, , \qquad
B \equiv 1+A
\, .
\eae
It will prove convenient in what follows to use the notation
$\lbbr  x \rbbr  \equiv \frac{x}{\{ x \} }$, with $\lbbr 0 \rbbr
\equiv 1$. Notice that $\lim_{q \rightarrow 1} \lbbr x \rbbr =
1$. We rewrite~(\ref{hparB})
\fle{hparBd}{%
\hpar =\bBb^{-1} \partial
\, .%
}
$\hpar$ is of the form $\hpar = f(B) \partial$, $f(B)
\equiv \lbbr B \rbbr^{-1}$. We look
for $\hx$ in the form $\hx = x g(B)$. Then
\bae
\label{hparhxcr}
\hpar \hx \fe f(B) \partial x g(B) \ff
 \fe f(B) g(B) + xf(B+1) g(B+1) \partial
\, .
\eae
The r.h.s. above should be equal to $1 + \hx \hpar$. One
concludes that $g(B)=f(B)^{-1}$, \ie,
\fle{hxdef}{%
\hx = x \bBb = \lbbr A \rbbr x
\, .%
}
The action of $\hx$ on monomials is
\be
\label{hxactxn}
\hx \tr x^n = \lbbr n+1 \rbbr x^{n+1}
\, .
\ee
$\hpar$ above acts on power series as a discrete derivative. We examine
the corresponding interpretation of the action of $\hx$. To this
end, we introduce the Jackson integral operator $S$ (see, \eg,
\cite{Ext:83}) given by
\be
\label{Sdef}
S \equiv \{ A \}^{-1} x 
\, .
\ee
Notice that $\{ A \}$ is invertible on the image of $x$,
\ie, \ on $x^n$, $n=1,2,\ldots$. Comparison with~(\ref{hparpseudo})
shows that $\hpar S=1$ while $(S\hpar) \tr x^n = x^n$,
$n=1,2,\ldots$ and $(S\hpar) \tr 1 = 0$. 

\par
\setlength{\figurewidth}{.4\textwidth}
\begin{floatingfigure}{\figurewidth}
\rule{0mm}{.675\figurewidth}
\begin{pspicture}(0\figurewidth,-.1\figurewidth)%
                 (\figurewidth,.7\figurewidth)
\setlength{\unitlength}{.25\figurewidth}
\psset{xunit=.25\figurewidth,yunit=.25\figurewidth,arrowsize=1.5pt 3} 
\put(.73,-.24){\makebox[0cm][c]{$q^3x$}}
\put(1.133,-.24){\makebox[0cm][c]{$q^2x$}}
\put(1.6875,-.24){\makebox[0cm][c]{$qx$}}
\put(2.53,-.24){\makebox[0cm][c]{$x$}}
\put(0,2.6){\makebox[0cm][c]{$f(x)$}}
\psline[linewidth=.3mm]{->}%
(0,0)(3.5,0)
\psline[linewidth=.3mm]{->}%
(0,0)(0,2.5)
\pscurve[linewidth=.3mm]{-}%
(.2,.5)(.5,.8)(.75,1.2)(1.125,1.7)(1.6875,2)(2.53,1.5)(3.1,1.1)

\psline[linewidth=.2mm,linestyle=dashed]{-*}%
(.5,1.2)(.5,0)
\psline[linewidth=.2mm,linestyle=dashed]{-*}%
(.75,1.7)(.75,0)
\psline[linewidth=.2mm,linestyle=dashed]{-*}%
(1.125,2)(1.125,0)
\psline[linewidth=.2mm,linestyle=dashed]{-*}%
(1.6875,2)(1.6875,0)
\psline[linewidth=.2mm,linestyle=dashed]{-*}%
(2.53,1.5)(2.53,0)

\psline[linewidth=.2mm,linestyle=dashed]{-}%
(.5,1.2)(.75,1.2)
\psline[linewidth=.2mm,linestyle=dashed]{-}%
(.75,1.7)(1.125,1.7)
\psline[linewidth=.2mm,linestyle=dashed]{-}%
(1.125,2)(1.6875,2)
\psline[linewidth=.2mm,linestyle=dashed]{-}%
(1.6875,1.5)(2.53,1.5)
\end{pspicture}%
\capitem{The action of $S$ on $f$, evaluated at $x$, gives the
area under the dotted lines.} 
\label{intfig}
\end{floatingfigure}
\par

Using the expansion
\be
\label{Sexpand}
 S = (1-q) \sum_{n=0}^{\infty} q^{nA} x \, ,
\ee
 one finds
\be
\label{Sactf}
 S \tr f(x) = (1-q) \sum_{n=0}^{\infty} q^n x f(q^n x) \, ,
\ee
 \ie, $S \tr f(x)$ gives the area under the dotted lines in
Fig.~\ref{intfig} and converges to $\int_{0}^{x} dx' f(x')$
in the limit $q \rightarrow 1$. Using the second of~(\ref{hxdef}),
we find
\fle{hxxpS}{%
\hx = x \partial S
\, .%
}%
In other words, the action of $\hx$ on $f(x)$ consists in
first producing the function $\tilde{f}(x) \equiv
(\partial S) \tr f(x)$ and then multiplying this latter
by the classical coordinate $x$.

\begin{aside}
We derive an alternative
expression for $\tilde{f}(x)$. In classical calculus one
has (Rolle theorem)
\be
\label{classf} f(x) = \averx{f} + \averx{xf'} \, , \ee where
$\averx{\cdot}$ denotes (classical) averaging in the interval
$[0,x]$, $\averx{f} \equiv \frac{1}{x} \int_0^x dx' f(x')$  and
$f'(x)$ is the (classical) derivative w.r.t. $x$.
Let now
$\averqx{\cdot}$ denote
{\em quantum averaging},
\be
\label{averqxdef}
\averqx{f} \equiv \frac{1}{x} S \tr f(x)
\, .
\ee
Then
\bae
(\partial S) \tr f(x)
\fe
\partial (1-q) \sum_{n=0}^\infty q^n x f(q^n x)
\ff \fe (1-q) \sum_{n=0}^\infty q^n
  \bigl(f(q^n x) + x q^n f'(q^n x) \bigr)
\ff \Ra \, \, \tilde{f}(x) \fe \averqx{f} + \averqx{xf'} \, ,
\eae
which is the $q$-deformed (``quantum'') analogue of~(\ref{classf}).
\end{aside}
Since $\hpar$, $\hx$ obey the CCR, one can define a {\em
quantum action} $\tr_q$, in complete analogy to the classical
one,
\be
\label{qaction}
\hpar \tr_q f(\hx) \equiv \hpar f(\hx) \vacz_{\hx,\hpar}
\, ,
\ee
where, in the r.h.s., $\hpar$ is commuted past $\hx$ until it
reaches the vacuum, where it gets annihilated -- the remaining
function of $\hx$ is, by definition, $\hpar \tr f(\hx)$ (notice
that the subscript of $\vacz$ instructs to express everything in
terms of $\hx$, $\hpar$).
This extends to arbitrary operators $G(\hx, \, \hpar)$
acting on functions of $\hx$, just like in the classical
case.  It follows trivially that\footnote{We have
$\phi_q\bigl(G(x,\partial)\bigr) = G(\hx,\hpar)$ --- the ordering of
the $x$'s and $\partial$'s in $G(x,\partial)$ is immaterial, as
long as the same ordering is used in $G(\hx,\hpar)$.}
\be
\label{qactphi}
 G(\hx, \, \hpar) \tr_q f(\hx) = \phi_q \bigl( G(x,
 \, \partial) \tr f(x) \bigr) \, ,
\ee
where $\phi_q:\, x\mapsto x_q$, $\partial \mapsto \partial_q$, is
the $q$-deformation map.
We note in passing that the quantum averaging operator,
$M_q \equiv \frac{1}{x} S $, is the inverse of $B$. Notice also
that the product $\hx \hpar $ is constant in $q$ (\ie,
invariant under the deformation),
\be
\hx \hpar \, = \, x \partial \, = \, A.
\ee
This implies that, when dealing with the $q$-deformation of a general
differential operator, terms of the form $x_q^m \partial_q^n$,
with $m \leq n$, can be expressed entirely in terms of $A$ and
$\partial_q$, the action of which is simpler. The resulting
$q$-deformed operator is a differential-difference operator.
This occurs, for example, in the study of the hypergeometric
operator.
Notice also that the invariance of $A$ permits the exponentiation
of the infinitesimal generator of $\phi_q$ in a trivial manner.
Finally, it is worth pointing out that the map $\phi_q$ admits a
non-trivial classical limit, which preserves Poisson brackets.
Indeed, with $p$, $q$ satisfying $\{ p, \, q \} =1$, where $\{
\cdot, \, \cdot \}$ is the Poisson bracket, one can easily verify
that $\{ f(A) p, \, q f(A)^{-1} \} =1$, where $A \equiv q p$,
giving rise to a wide class of classical canonical transformations
that might in itself be worth exploring.
\section{CCR-preserving Maps and Adapted Bases}
\label{CCRpMAB}
Consider the Heisenberg-Weyl universal enveloping algebra $\calH$,
generated by $a$, $b$, with $[a,\, b]=1$. In the sequel we work
with a certain completion $\hH$ of $\calH$, that allows us to deal
with \eg, exponentials in $a$, $b$. From the discussion of the
previous section (see~(\ref{qactphi})), we abstract  a map
$\phi_q : \hH \rightarrow \hH$ that preserves the CCR%
\footnote{%
We use the same symbol as for the map in~(\ref{qactphi})%
}
\be
\label{phiqdef} 
\phi_q : \, a \mapsto a_q \equiv \bBb^{-1} a 
\, ,
\qquad 
b \mapsto b_q \equiv b \bBb 
\, , 
\ee 
where $B$ now is $B = 1 + ba$. 
All equations in the previous section depend only on $x$,
$\partial$ satisfying the CCR and are therefore valid for
$(x,\, \partial) \mapsto (b, \, a)$. Although we use the
particular map $\phi_q$ given above as an example, we emphasize
that our results below are general.

For any pair $(a,b)$ of abstract generators that satisfy the
CCR, we say that the set $\{ \ketn , \, n=0,1,2,\ldots \}$
is an {\em adapted basis for $(a,b)$} if $a$, $b$ act on it as
lowering and raising operators respectively
\be
\label{abadapt}
a \ketn = n \ketnm
\, ,
\qquad
\qquad
\qquad
b \ketn = \ketnp
\, .
\ee
\begin{example}{A classical adapted basis}{cab}
One particular representation of the Heisenberg algebra
is supplied by the subalgebra generated by $\{1,b\}$, an
adapted basis being given by $\ketn =b^n$. The action of
the Heisenberg algebra generators on an arbitrary power
series $f(b)$ is $a \tr f \equiv [a,f]$ and $b \tr f
\equiv bf$.
\end{example}

\noindent Suppose now we are given a CCR-preserving map
$\phi_\alpha$, where $\alpha$ denotes any parameters $\phi$ might
depend on, and we wish to find an adapted basis for the
deformed generators it produces. A general
solution to this problem is possible if we further impose the
restriction that $\phi_\alpha$ be {\em counit preserving} (\ie,
$\phi_\alpha(a) \vacz=0$). It is worth emphasizing that this 
requirement, although rather natural, excludes nevertheless a number 
of familiar
CCR-preserving maps, like the rotation from $x$, $\partial$ to
$a^\dagger$, $a$ in the simple harmonic oscillator. Keeping this
observation in mind, we proceed to the following statement: 
given any CCR {\em and} counit-preserving
map $\phi_\alpha$,
one can
find in general an {\em induced} map $\tilde{\phi}_\alpha: \, \ketn
\mapsto \ketan$ that maps {\em any} adapted basis $\{
\ketn \}$ for $(a,b)$ to an adapted basis $\{ \ketan \}$
for $(a_\alpha, b_\alpha)$.  
Indeed, to any function $f(b_\alpha)$ one can
associate its $b$-{\em projection} $\tf(b)$ given
by%
\footnote{%
Notice that the
$b$-projection of $f$ depends on the particular deformation
$\phi_\alpha$ used --- for simplicity of notation we do not
show this dependence explicitly.%
}
\be
\label{bprojectdef}
\tf(b) = f(b_\alpha) \vacz_{b,a}
\, .
\ee
We now show that
\fle{aqftb}{%
G_\alpha \tr \tf = \widehat{G \tr f}
\, ,%
}%
where $f=f(b)$, $G=G(b, \, a)$ and $G_\alpha \equiv
\phi_\alpha\bigl(G(b,\, a)\bigr) = G(b_\alpha, \, a_\alpha)$. We
have 
\bae 
G_\alpha \tr \tf (b) 
& \equalref{actdef} & 
G_\alpha \tf (b) \vacz_{b,a} 
\ff 
& \equalref{bprojectdef} & 
G_\alpha f(b_\alpha) \vacz_{b,a} 
\ff 
& \equalref{qaction} & 
G_\alpha \tr_\alpha f(b_\alpha) \vacz_{b,a} 
\ff 
& \equalref{qactphi} &
\phi_\alpha \left( G \tr f(b) \right) \vacz_{b,a} 
\ff 
\fe 
(G \tr f)(b_\alpha) \vacz_{b,a} 
\ff 
& \equalref{bprojectdef} & 
\widehat{G \tr f}(b) 
\, . 
\label{aqfbpr} 
\eae 
We comment briefly on the steps
that lead to~(\ref{aqfbpr}). The first equality follows
from~(\ref{actdef}), taking into account that $\phi_\alpha$ is
counit preserving, so that we may put $a$ in place of $a_\alpha$
in the subscript of the vacuum. The second equality follows
from~(\ref{bprojectdef}). In the expression $G_\alpha f(b_\alpha)
\vacz_{b,a}$ we are instructed to express $a_\alpha$, $b_\alpha$
in terms of $a$, $b$, and then bring the $a$'s to the right \etc..
One can do this though in several ways. The one shown above
involves first bringing all $a_\alpha$'s to the right of the
$b_\alpha$'s, then substituting $a_\alpha= a_\alpha(b,a)$ and
bringing the $a$'s to the right (this is equivalent to
annihilating the $a_\alpha$'s themselves, since $\phi_\alpha$
preserves the counit). At this point one is left with a function
of $b_\alpha$ which is clearly $G_\alpha \tr_\alpha f(b_\alpha)$.
Finally, one substitutes $b_\alpha = b_\alpha (b,a)$ and brings
the $a$'s to the right.

Given an adapted basis $\{ \ketn \}$ for $(a,b)$,
we construct the set $\{ \ketan \}$, where\footnote{%
The vacuum
used in computing $\widehat{b^n} \equiv b_\alpha^n \vacz$ is the
ket $| n=0 \rangle$.}
\fle{ketqndef}{%
\ketan  \equiv \widehat{b^n}%
}%
and claim that it is an adapted basis for $(a_\alpha,b_\alpha)$.
Indeed,
\be
a_\alpha \ketan
\, \,
\equalref{ketqndef}
\, \,
a_\alpha \tr \widehat{b^n}
\, \,
\equalref{aqfbpr}
\, \,
\widehat{a \tr b^n}
\, \, = \, \,
n \widehat{b^{n-1}}
\, \, = \, \,
n \ketnm_\alpha
\, .
\ee
Also,
\be
\label{bqnqpr}
b_\alpha \ketan
\, \, = \, \,
b_\alpha \, b_\alpha^n \vacz
\, \, = \, \,
b_\alpha^{n+1} \vacz
\, \, = \, \,
\ketnp_\alpha
\, .
\ee
\begin{example}{A quantum adapted basis}{qabasis}
Continuing our earlier classical example, we now turn to the
realization of the Heisenberg algebra provided by the map $\phi_q$.
We take $\ketn =b^n$ and find for $\tilde{\phi}_q(\ketn) \equiv
\ketqn$
\bae
\label{nq}
\ketqn  \fe
\widehat{b^n}
\ff \fe
b_q^n \vacz
\ff \fe
( b \bBb )^n \vacz
\ff \fe
\bnb! \ketn
\, ,
\eae
where $\bnb! \equiv \lbbr 1 \rbbr \, \lbbr 2 \rbbr \dots \bnb$
and $\lbbr 0 \rbbr! \equiv 1$.
\end{example}
\begin{example}{A second discrete realization}{fdreal}
Consider the pair of operators
\be
\label{adbddef} 
a_\delta \equiv \phi_\delta(a) =  \delta^{-1} (e^{\delta a} -1) 
\, , 
\qquad \qquad 
b_\delta \equiv \phi_\delta(b) = b e^{-\delta a} 
\, . 
\ee
 One easily verifies that
$[a_\delta, \, b_\delta] =1$ --- this is the $\delta$-realization
of the CCR mentioned in the introduction. We take again $\ketn =
b^n$ and compute $\tilde{\phi}_\delta (\ketn)\equiv \ketn_\delta$
\bae 
\label{ketndel}
\ketn_\delta \fe b_\delta^n \vacz 
\ff 
& \equalref{adbddef} & 
b  e^{-\delta a} b e^{-\delta a} \dots be^{-\delta a} \vacz 
\ff 
\fe 
b(b-\delta)\cdot \dots \cdot \bigl(b-(n-1) \delta \bigr) 
e^{-n\delta a} \vacz 
\ff 
\fe
b(b-\delta)\cdot \dots \cdot \bigl(b-(n-1)\delta\bigr) 
\ff 
& \equiv & 
b_\delta^{(n)} 
\, .
\eae
For the {\em falling $\delta$-factorial polynomials}
$b_\delta^{(n)}$, defined by the next-to-last line above (also
known as {\em $\delta$-quasi-monomials}~\cite{Mil:51}), it
holds
\ble{dpoly}
b_\delta^{(n)} =\sum_{k=1}^n s(n,k) \delta^{n-k} b^k 
\, ,
\ee
where $s(n,k)$ are the Stirling numbers of the first kind.
\end{example}
\begin{example}{The $q$-exponential as a $b$-projection}{qebp}
Consider the spectral problem 
\[
\hpar \tr f(x) = \lambda f(x)
\, .
\]
Relying on~(\ref{aqftb}), we look instead at the equation
$\partial \tr g(x) = \lambda g(x)$ and compute $f(x)$ above from
$f = \widehat{g}$. We get $g \sim e^{\lambda x}$ and, using
$\widehat{x^n} \equalref{nq} \lbbr n \rbbr x^n$, we find
\bae
f(x)
& = & 
\widehat{g}(x) \ff \fe e^{\lambda \hx} \vacz_{x,\partial}
\ff 
\fe 
\sum_{n=0}^\infty \frac{\lambda^n}{n!} \hx^n \vacz_{x,\partial} 
\ff 
\fe 
\sum_{n=0}^\infty \frac{\lambda^n}{n!} \frac{n!}{\{ n \} !} x^n 
\ff 
\fe 
\sum_{n=0}^\infty \frac{1}{\{ n\} !} (\lambda x)^n 
\ff 
& \equiv & 
e_q(\lambda x) 
\, ,
\eae
\ie, the standard $q$-deformed exponential~\cite{Ext:83} is just
$e_q(x)=\widehat{e^x}$. More generally, if 
\[
a_\alpha = f_\alpha^{-1}(B) a
\, ,
\qquad \qquad
b_\alpha=b f_\alpha(B)
\, ,
\]
the eigenfunctions of $a_\alpha$ are $\sum_{n=0}^\infty
\frac{f_\alpha(n)!}{n!} b^n$, where $f_\alpha(n)! \equiv
f_\alpha(1)f_\alpha(2) \cdots f_\alpha(n)$, $f_\alpha(0)!
\equiv 1$.
\end{example}

\noindent It is important to emphasize at this point the
formal character
of the above results. In particular, a given eigenfunction $g(x)$
of some differential operator may converge for all $x$, while
its $b$-projection $\hat{g}(x)$ might have a finite (or even
zero) radius of convergence as it happens, for example, with the 
$\delta$-exponential 
\ble{dexp}
e_\delta(x) \equiv \widehat{e^{\lambda x}} = \sum_{n=0}^\infty
\frac{1}{n!} x_\delta^{(n)}
\ee
that appears as the projection
$\widehat{e^{\lambda x}}$ for the map $\phi_\delta$ of
Ex.~\ref{fdreal}. 
\section{Composition of CCR-preserving Maps}
\label{Compo}
If $\phi_\alpha$, $\phi_\beta$ are CCR-preserving maps then so is
their composition $\phi_\alpha \circ \phi_\beta$. Considering only
smooth maps, with a smooth inverse, one arrives at the notion of
the group of
CCR-preserving maps (CCR-PM). In the sequel we impose further the
requirement that our maps preserve the counit. We can then use
the fact that
\be
\widetilde{\phi_\alpha \circ \phi_\beta} = \tilde{\phi_\alpha}
\circ \tilde{\phi_\beta}
\label{comptilde}
\ee
to compute the induced map of a composition of maps. We
illustrate this in the following
\begin{example}{Composition of $\phi_q$, $\phi_\delta$}{cpqpd}
Consider the map $\phi_q$ discussed earlier. For
each value of $q$, $\phi_q$ is an element of the CCR-PM. However,
$\{\phi_q, \, -1 < q < 1\}$ is not a one-parameter subgroup since,
in general, there is no $q$ such that $\phi_{q_1} \circ
\phi_{q_2}=\phi_{q}$. Notice also that $[\phi_{q_1}, \,
\phi_{q_2} ] \neq 0$, for finite $q_1$, $q_2$. Similar remarks
hold for $\phi_\delta$. For the composition  $\phi_q
\circ \phi_\delta$ we find
\bae
\label{phiqdel}
a_{q \delta}
& \equiv &
(\phi_q \circ \phi_\delta) (a)
\ff \fe
\phi_q(\delta^{-1} (e^{\delta a} -1))
\ff \fe
 \delta^{-1} \left( e^{\delta \bBb^{-1} a} -1 \right)
\, ,
\ff
& & \ff
b_{q\delta}
& \equiv &
(\phi_q \circ \phi_\delta) (b)
\ff \fe
\phi_q(be^{-\delta a})
\ff \fe
b \lbbr B \rbbr e^{-\delta \bBb^{-1} a}
\, ,
\eae
while for $\phi_\delta \circ \phi_q$ we get
\bae
\label{phidelq}
a_{\delta q}
& \equiv &
(\phi_\delta \circ \phi_q) (a)
\ff \fe
\phi_\delta ( \bBb^{-1} a )
\ff \fe
\delta^{-1}
\lbbr  1+\delta^{-1} b (1-e^{-\delta a}) \rbbr^{-1}
(e^{\delta a } -1)
\, ,
\ff
& &  \ff
b_{\delta q}
& \equiv &
(\phi_\delta \circ \phi_q) (b)
\ff \fe
\phi_\delta ( b \bBb )
\ff \fe
b e^{-\delta a }
\lbbr 1+\delta^{-1}b(1-e^{-\delta a}) \rbbr
\, .
\eae
For the adapted bases that correspond to the above compositions
we find
\bae
\label{adbphiqd}
\ketn_{q\delta} & \equiv & \widetilde{\phi_q \circ \phi_\delta}
(\ketn)
\ff \fe
\phi_q(b_\delta^{(n)})
\ff \fe
\sum_{k=1}^n s(n,k) \delta^{n-k} \phi_q(b^k)
\ff \fe
\sum_{k=1}^n s(n,k) \delta^{n-k} \lbbr k \rbbr b^k
\, .
\eae
For example,
\be
\label{ex2qd}
|2 \rangle_{q\delta}
= \phi_q\left(b(b-\delta)\right)
= \lbbr 2 \rbbr b^2 -\delta b
= \frac{2}{1+q} b^2 -\delta b
\, .
\ee
Also,
\bae
\label{adbphidq}
\ketn_{\delta q} & \equiv & \widetilde{\phi_\delta \circ \phi_q}
(\ketn)
\ff \fe
\tilde{\phi}_\delta \left(\lbbr n \rbbr ! \ketn \right)
\ff \fe
\lbbr n \rbbr! \, b_\delta^{(n)}
\, ,
\eae
so that, for example,
\be
\label{ex2dq}
| 2 \rangle_{\delta q} = \frac{2}{1+q} b(b-\delta)
\, ,
\ee
which should be compared with~(\ref{ex2qd}).
\end{example}
\section{Quantum Canonical Conjugates}
\label{qCC}
Given $x$, $y$ satisfying the $q$-Heisenberg algebra
\be
\label{qHa}
 xy -qyx=1 \, ,
\ee
with $-1 < q < 1$. We say that $x$ is the {\em quantum canonical
conjugate} (QCC) of $y$ (and {\em vice versa}). To complete our
treatment of the map $\phi_\delta$ of Ex.~\ref{fdreal}, we
undertake here the determination of the QCC of $a_\delta$. We work
again with abstract operators $a$, $b$ and remark that $a_q$
(given in~(\ref{phiqdef})) and $b$ satisfy~(\ref{qHa}), $a_q b - q
b a_q=1$ . Notice that, the (classical) Heisenberg algebra admits
the $*$-involution
\be
\label{starinv}
b^* = a \, , \qquad \qquad a^* =b 
\, ,
\ee
which we extend as complex conjugation to the parameter $q$, 
$q^* = q$. Then, taking the $*$ of~(\ref{qHa}) (with $a_q$
expressed in terms of $a$, $b$, as in~(\ref{hparBd})), we find
\bae 
\label{bsas} 
b^* {a_q}^* -q {a_q}^* b^*  
\fe 1 
\ff
\Rightarrow 
\, \, a \, 
\left(b \lbbr B \rbbr^{-1} \right) 
-q \left( b \lbbr B \rbbr^{-1} \right)\, a 
\fe 1 
\, . 
\eae 
Up to now
we disposed of the deforming map $ \phi_q: \, a \mapsto a_q$, $b
\mapsto b$, which, applied to a pair of operators satisfying
the classical Heisenberg algebra, produces a pair satisfying the
quantum Heisenberg algebra. Notice that it does so by leaving $b$
invariant and only deforming $a$. What we have achieved
in~(\ref{bsas}), is to produce a second similar map $\phi'_q$,
which instead leaves $a$ invariant and deforms
only $b$: $\phi'_q: \, a \mapsto a$, $b \mapsto {a_q}^*$. We only need
apply $\phi_\delta$ to~(\ref{bsas}) to get
\be
\label{adqcc}
a_\delta \, \phi_\delta({a_q}^*) -q \phi_\delta({a_q}^*) \,
a_\delta =1
\, ,
\ee
which identifies
$\phi_\delta({a_q}^*)=\phi_\delta\left( b \lbbr B \rbbr^{-1}
\right)$ as the QCC of $a_\delta$.
\section{Deformed Hahn Polynomials}
\label{Hahn}
As an example of potential applications of our results, we
present here various deformations of the Hahn operator and its
eigenfunctions, the Hahn polynomials. We start with
some definitions. The action of the {\em Hahn operator} $H_\delta$ on 
a function $f(x)$ is given by
\bae
\label{hahn}
H_\delta \tr f(x) 
& \equiv &
\delta^{-3} \bigl( c_4 \delta^2 + c_2 \delta x + c_1 x^2 \bigr)
\, f(x+\delta) 
\ff
 & & 
{} - \delta^{-3} \bigl( c_4 \delta^2 
- \delta(c_1 - 2 c_2 + c_3 \delta)x + 2 c_1 x^2 \bigr) \, f(x) 
\ff
& & 
{} - \delta^{-3} 
\bigl( \delta(c_1 - c_2 + c_3 \delta )x - c_1 x^2 \bigr) 
\, f(x-\delta) 
\, ,
\eae
where $c_i$, $\delta$ are parameters.
What distinguishes $H_\delta$ is
that it is the most general three-point finite-difference
operator with infinitely many polynomial
eigenfunctions~\cite{Smi.Tur:95}. The  latter are called {\em
Hahn polynomials} (of continuous argument) and we denote them by 
$h_k^{(\alpha,\beta;\delta)}(x, N)$, where
\ble{hahn-par}
c_2 = N - 2 - \beta
\, ,
\qquad \qquad
c_3 = -\alpha-\beta-1
\, ,
c_4=(\beta+1)(N-1) 
\, ,
\ee
and we have set, without loss of generality, $\delta=1$ and
$c_1=-1$. The corresponding eigenvalues are
\ble{hahn-spec}
\lambda_k = \delta^{-1} c_1 k^2 + c_3 k
\, ,
\qquad
k=0,1,2,\ldots\ .
\ee
For particular values of their parameters and/or arguments,
$h_k^{(\alpha,\beta;\delta)}(x, N)$ reduce to the 
Meixner, Charlier, Tschebyschev, Krawtchouk or (discrete
argument) Hahn polynomials (for
details and their r\^ole in finite difference equations, 
see~\cite{Ash:85,Ata.Sus:85,Nik.Sus.Uva:91} and references
therein).
We will use the form
\ble{hahn-pol}
h_k^{(\alpha,\beta;\delta)} (x, N) 
= \sum_{i=0}^{k} \gamma_i x_\delta^{(i)}
\, ,
\ee
where $x^{(i)}$ is as in~(\ref{dpoly}) and the
$\gamma_i$ are known coefficients. 

It has been shown in~\cite{Smi.Tur:95, Smi.Tur:96} that
$H_\delta$ belongs to $\hat{\calH}$,
\ble{adbd-hahn}
H_{\delta} = 
c_1
(b_{\delta}a_{\delta})^2 ( a_{\delta}+\delta^{-1} )
 + c_2 b_{\delta} a_{\delta}^2 + c_3 b_{\delta} a_{\delta}
 + c_4 a_{\delta} 
\, ,
\ee
where $a_\delta$, $b_\delta$ are given in~(\ref{adbddef}), with
$(a,b) \mapsto (\partial,x)$. We are now at a setting where our 
earlier results may be applied directly. First, we deform
isospectrally $H_\delta$ to $H$, by effecting the substitution
$(a_\delta,b_\delta) \mapsto (\partial,x)$ in~(\ref{adbd-hahn})
--- a little bit of algebra gives
\ble{ab-hahn}
H = c_1 \, x^2 {\partial}^3 
+ [(c_1+c_2)+ c_1 \delta^{-1} x] \, x {\partial}^2 
+ [c_4 + (c_1 \delta^{-1} + c_3)x] \, \partial 
\, .
\ee
The corresponding polynomial eigenfunctions are obtained
directly from~(\ref{hahn-pol}), using the results of
Ex.~\ref{fdreal},
\ble{hahn-pol-dif}
{\tilde h}_k^{(\alpha,\beta)} (x, N) = \sum_{i=0}^{k} \gamma_i x^i
\ee
(the eigenvalues are, of course, still given
by~(\ref{hahn-spec})). A second isospectral deformation,
involving the substitution $(\partial,x) \mapsto (\hpar,\hx)$
in~(\ref{ab-hahn}), leads to
\ble{aqbq-hahn}
 H_{q} = c_1 A^2 (\hpar + \delta^{-1})
 + c_2 A \hpar 
 + c_3 A
 + c_4 \hpar 
\, ,
\ee
($A = x \partial$) with polynomial eigenfunctions
\ble{hahn-pol-q}
h_k^{(\alpha,\beta;q)} (x, N) 
= \sum_{i=0}^{k} \gamma_i \lbbr i \rbbr! \, x^i
\, .
\ee
Notice the ease with which $h_k^{(\alpha,\beta;q)} (x, N)$ are
computed, despite the  highly non-trivial complexity of the
differential-difference operator $H_q$. Finally, we point out
that the {\em spectrum} of $H$ in~(\ref{ab-hahn}) may also 
be $q$-deformed
by effecting the substitution $(\partial, x) \mapsto (\hpar,x)$
in~(\ref{ab-hahn}) --- one gets a finite-difference operator
\ble{aqb-hahn}
\tilde{H}_q =
c_1 \, x^2 {\partial_q}^3 
+ (c_1+c_2 + c_1  \delta^{-1} x) \, x {\partial_q}^2 
+ [c_4 + (c_1 \delta^{-1} + c_3)x] \, \partial_q 
\, ,
\ee
with infinitely many polynomial eigenfunctions --- the
corresponding eigenvalues are
\ble{hahn-spec-q}
\tilde{\lambda}_k 
= c_1 \delta^{-1} \{ k \} \bigl( \{ k-1 \} + 1 \bigr) + c_3 \{ k \}
\, ,
\qquad 
k=0,1,2,\ldots
\, .
\ee

\vspace{2mm}

\noindent \textbf{Note Added} \vspace{1mm}\\
After this work was sent for publication, Professor C.~Zachos,
to whom we express our gratitude, 
pointed out to us that $\hx$, $\hpar$ can be  related to $x$,
$\partial$ via a similarity transformation. Indeed, starting with the
ansatz $U(A)^{-1} \partial U(A) = \hpar$, it follows that
$\partial U(A) = \hpar U(A-1)$ and, using~(\ref{hparpseudo}),
$U(A)= \lbbr A \rbbr^{-1} U(A-1)$, from which the formal
expression $U(A)=\Gamma_q(A+1)/\Gamma(A+1)$ can be derived. Here
$\Gamma_q$ denotes the $q$-deformed gamma function, 
$\Gamma_q(x+1) = \{ x \} \Gamma_q(x)$ (see, \eg,~\cite{Ext:83}).
Then, $\hx$ is computed as $\hx = U(A)^{-1} x U(A) =
U(A)^{-1} U(A-1) x = \lbbr A \rbbr x$ and the constancy of $A$ in $q$
follows trivially.
 

\begin{thebibliography}{10}

\bibitem{Ana.Bow.Sch:92}
K.{}~N.{} Anagnostopoulos, M.{}~J.{} Bowick, and A.{} Schwarz.
\newblock The {S}olution {S}pace of {U}nitary {M}atrix {M}odel {S}tring
  {E}quation and the {S}ato {G}rassmannian.
\newblock {\em Commun.{} Math.{} Phys.{}}, 148:469--486, 1992.

\bibitem{Ash:85}
R.~Askey.
\newblock Continuous {H}ahn {P}olynomials.
\newblock {\em Journ. Phys.}, A18:L1017--L1019, 1985.

\bibitem{Ata.Sus:85}
N.M. Atakishiev and S.K. Suslov.
\newblock Hahn and {M}eixner {P}olynomials of an 
{I}maginary {A}rgument and some of their {A}pplications.
\newblock {\em Journ. Phys.}, A18:1583--1596, 1985.

\bibitem{Ext:83}
H.~Exton.
\newblock {\em q-hypergeometric {F}unctions and {A}pplications}.
\newblock Ellis Horwood, 1983.

\bibitem{Gor.Szm:98}
A.{}~Z.{} G\'orski and J.{} Szmigielski.
\newblock On {P}airs of {D}ifference {O}perators {S}atisfying
  $[d,x]=\mathrm{id}$.
\newblock {\em Jour.{} Math.{} Phys.{}}, 39:545--568, 1998.

\bibitem{Gor.Szm:00}
A.{}~Z.{} G\'orski and J.{} Szmigielski.
\newblock Representations of the {H}eisenberg {A}lgebra by {D}ifference
  {O}perators.
\newblock {\em Acta Phys.{} Polon.{}}, B31:789--799, 2000.

\bibitem{Mil:51}
L.{}~M.{} Milne-Thomson.
\newblock {\em The {C}alculus of {F}inite {D}ifferences}.
\newblock MacMillan and Co. Limited, London, 1951.

\bibitem{Nik.Sus.Uva:91}
A.{}~F.{} Nikiforov, S.{}~K.{} Suslov, and V.{}~B.{} Uvarov.
\newblock {\em Classical {O}rthogonal {P}olynomials of a {D}iscrete
  {V}ariable}.
\newblock Springer Series in Computational Physics. 
Springer-Verlag, 1991.

\bibitem{Sch:91}
A.~Schwarz.
\newblock On {S}olutions to the {S}tring {E}quation.
\newblock {\em Mod. Phys. Lett.}, A6:2713--2726, 1991.

\bibitem{Smi.Tur:95}
Yu.{}~F.{} Smirnov and A.{} Turbiner.
\newblock Lie-algebraic {D}iscretization of {D}ifferential {E}quations.
\newblock {\em Mod.{} Phys.{} Lett.{}}, A10:1795--1802, 1995.
\newblock Erratum, {\em ibid} A10:3139, 1995.

\bibitem{Smi.Tur:96}
Yu.{}~F.{} Smirnov and A.{} Turbiner.
\newblock Hidden $sl_2$ {A}lgebra of {F}inite-difference {E}quations.
\newblock In N.{}~M.{} Atakishiyev, 
T.{}~H.{} Seligman, and K.{}~B.{} Wolf,
  editors, {\em Proceedings of the {IV} {W}igner {S}ymposium}. World
  Scientific, 1996.

\bibitem{Tur:00}
A.{} Turbiner.
\newblock Canonical {D}iscretization. {I}. {D}iscrete {F}aces of 
({A}n)harmonic {O}scillator.
\newblock {\em Intern.{} Jour.{} Mod.{} Phys.{}}, A16:1579--1605, 2001.
\newblock \texttt{hep-th/0004175}.

\end{thebibliography}

\end{document}